\documentclass[conference]{IEEEtran}
\IEEEoverridecommandlockouts
\usepackage{cite}
\usepackage{amsmath,amssymb,amsfonts}
\usepackage{algorithmic}
\usepackage{graphicx}
\usepackage{textcomp}
\usepackage{comment}
\usepackage{kotex}
\usepackage{xcolor}
\usepackage{xcolor,colortbl}
\usepackage{amsmath,amssymb,amsfonts}
\usepackage{comment}
\usepackage{algorithmic}
\usepackage{graphicx}
\usepackage{mathtools}
\usepackage{textcomp}
\usepackage{setspace}
\usepackage{pifont}
\usepackage{soul}
\usepackage{siunitx}
\usepackage{wrapfig}
\usepackage[normalem]{ulem}
\usepackage{amsfonts}
\usepackage{microtype}
\usepackage{textcase}
\usepackage{multirow}
\usepackage{caption}
\usepackage{subcaption}
\usepackage[colorlinks=true,linkcolor=purple,citecolor=purple,urlcolor=black]{hyperref}
\newcolumntype{L}{>{\raggedright\arraybackslash}X}
\usepackage{textcomp}
\usepackage{tcolorbox}

\usepackage[linesnumbered,ruled,vlined]{algorithm2e}
\usepackage{titlecaps}
\usepackage{url,tikz}

\newcommand{\bcircled}[1]{\tikz[baseline=(char.base)]{ \node[shape=circle,fill,inner sep=1pt] (char) {\textcolor{white}{#1}};}}

\newcommand{\llc}{L\textsubscript{0} to L\textsubscript{1} compaction}
\newcommand{\lz}{L\textsubscript{0}}

\newcommand{\squishlist}{
\begin{list}{$\bullet$}
	{ \setlength{\itemsep}{0pt}      \setlength{\parsep}{-0pt}
		\setlength{\topsep}{4pt}       \setlength{\partopsep}{0pt}
		\setlength{\listparindent}{-2pt}
		\setlength{\itemindent}{-5pt}
		\setlength{\leftmargin}{1em} \setlength{\labelwidth}{0em}
		\setlength{\labelsep}{0.5em} } }
  
\newcommand{\squishend}{
\end{list}  }
\def\BibTeX{{\rm B\kern-.05em{\sc i\kern-.025em b}\kern-.08em
    T\kern-.1667em\lower.7ex\hbox{E}\kern-.125emX}}
\begin{document}

\title{\huge
A Host-SSD Collaborative Write Accelerator for LSM-Tree-Based Key-Value Stores
}

\newcommand{\kcache}{{\titlecap{\scshape KVAccel}\normalfont}}

\author{\IEEEauthorblockN{Kihwan Kim$^{1,*}$,
Hyunsun Chung$^{1,*}$, Seonghoon Ahn$^{1,*}$, Junhyeok Park$^1$, Safdar Jamil$^1$\\
Hongsu Byun$^1$, Myungcheol Lee$^2$, Jinchun Choi$^2$, Youngjae Kim$^{1,\dagger{}}$
\thanks{$^{*}$They are first co-authors and have contributed equally.}
\thanks{$^{\dagger}$Y. Kim is the corresponding author.}
}
\IEEEauthorblockA{$^1$Sogang University, Seoul, Republic of Korea, $^2$ETRI, Daejeon, Republic of Korea} 
}

\maketitle
\thispagestyle{plain}
\pagestyle{plain}

\setstretch{0.988}

\begin{abstract}
Log-Structured Merge (LSM) tree-based Key-Value Stores (KVSs) are widely adopted for their high performance in write-intensive environments, but they often face performance degradation due to write stalls during compaction. Prior solutions, such as regulating I/O traffic or using multiple compaction threads, can cause unexpected drops in throughput or increase host CPU usage, while hardware-based approaches using FPGA, GPU, and DPU aimed at reducing compaction duration introduce additional hardware costs. In this study, we propose \kcache{}, a novel hardware-software co-design framework that eliminates write stalls by leveraging a dual-interface SSD. \kcache{} allocates logical NAND flash space to support both block and key-value interfaces, using the key-value interface as a temporary write buffer during write stalls. This strategy significantly reduces write stalls, optimizes resource usage, and ensures consistency between the host and device by implementing an in-device LSM-based write buffer with an iterator-based range scan mechanism. Our extensive evaluation shows that for write-intensive workloads, \kcache{} outperforms ADOC by up to 17\% in terms of throughput and performance-to-CPU-utilization efficiency. For mixed read-write workloads, both demonstrate comparable performance.
\end{abstract}

\section{Introduction}
\label{sec:intro}

Log-Structured Merge (LSM) tree-based Key-Value Store (KVS) systems, such as RocksDB~\cite{rocksdb} and LevelDB~\cite{levelDB}, are commonly used in write-intensive applications due to their ability to handle high-throughput writes efficiently. However, LSM-based KVSs (LSM-KVSs) often experience performance degradation due to write stalls that occur during compaction~\cite{silk, 10027097, yu2023adoc, yao2020matrixkv, balmau2017triad, ding2022trianglekv}. These write stalls block incoming write operations, resulting in a significant reduction in throughput and an increase in tail latency, which undermines system reliability in time-sensitive workloads.

To alleviate write stalls, RocksDB~\cite{rocksdb}, one of the most widely used LSM-KVS, implements a mechanism known as \textit{slowdown}~\cite{rocksdb_slowdown}. This slowdown mechanism anticipates potential write stalls and proactively reduces the write pressure on the LSM-KVS. 
Consequently, while it can prevent write stalls, it may unnecessarily decrease the throughput of RocksDB by limiting the write pressure directed to the LSM-KVS. 
Additionally, the state-of-the-art solution ADOC~\cite{yu2023adoc} mitigates write stalls by dynamically increasing batch sizes and the number of compaction threads during a write slowdown, thereby reducing compaction duration.
However, ADOC increases host CPU utilization by employing multiple compaction threads.

Alternatively, hardware-based solutions have been investigated. Persistent memory (PM)-based designs~\cite{yao2020matrixkv, kannan2018redesigning, kaiyrakhmet2019slm} buffer writes in PM before flushing them to the LSM-tree, while FPGA-based accelerators~\cite{sun2020fpga,fpga_compaction,huang2019x}, GPU~\cite{xu2020luda,zhougpu,sun2024glsm}, and DPU~\cite{ding2023dcomp,ding2024d}
speed up merge sort to reduce compaction time. Key-Value SSD (KV-SSD) architectures~\cite{KVSSDsamsung, jin2017kaml, lee2019ilsm, im2020pink, iteratorkvssd, bandslim} handle key-value operations directly within storage devices, bypassing the OS and file system overheads. Although these approaches enhance performance, they require additional hardware (e.g., PM, FPGA, GPU, DPU), raising costs and complexity.

The aforementioned software solutions suffer from unnecessary performance degradation due to inaccurate predictions or increased host CPU usage, while hardware solutions require additional hardware, raising costs. In this study, we propose a groundbreaking approach that avoids write stalls without compromising KVS performance, minimizes host CPU utilization, and requires no additional hardware costs. 
Our method represents a new paradigm, fundamentally different from existing approaches, by actively leveraging idle resources in existing storage devices to avoid write stalls while minimizing host CPU involvement.

In this paper, we present \kcache{}, a novel hybrid hardware-software co-design framework that leverages a new dual-interface SSD architecture to mitigate write stalls and optimize the utilization of storage bandwidth. 
\kcache{} is built on the observation that during host-side write stalls, the underlying storage device’s available I/O bandwidth remains underutilized, despite its potential to handle additional I/O operations.
\kcache{} then incorporates a dynamic I/O redirection mechanism that monitors the status of host-side LSM-KVS and, upon detecting a write stall, shifts writes from the LSM-KVS to the device-side key-value write buffer. 

\kcache{} presents a disaggregation of the SSD’s logical NAND flash address space into two regions: one for the traditional block interface, which is managed by the host-side LSM-KVS, and another for the key-value interface inspired by the KV-SSD, which serves as a temporary write buffer to serve pending write requests by bypassing the traditional LSM-based data path during stalls. 

To maintain consistency between the main LSM on the host and the write buffer on the device, \kcache{} introduces a range scan-based rollback mechanism. 
This mechanism structures the device-side write buffer as a separate LSM from the host-side main LSM and employs an iterator-based range scan over the buffer, enabling fast scan of buffered key-value pairs back to the host for merging. 
\kcache{} then merges them into the main LSM, maintaining the properties of the LSM and ensuring data consistency between the two interfaces.

\kcache{} offers detector, I/O redirection, and rollback modules on top of RocksDB~\cite{rocksdb}.
The dual-interface SSD was implemented using the Cosmos+ OpenSSD platform~\cite{cosmos_plus_open_ssd}, an FPGA-based NVMe SSD development board. 
RocksDB operates on the block interface provided by a single OpenSSD, while the key-value interface of the same device handles the redirected key-value pairs.

The key contributions of this paper are as follows:
\squishlist
\item We identify a critical opportunity to mitigate the fundamental issue of write stalls in LSM-KVS by leveraging the underutilized storage bandwidth during these stalls, transforming an inherent inefficiency into a performance optimization.

\item We propose a hybrid SSD architecture that integrates a new key-value interface alongside the traditional block interface within a singular device, allowing us to address write stalls without significantly modifying the existing LSM-KVS or deploying additional hardware in the system.

\item We develop efficient dynamic I/O redirection and rollback mechanisms to seamlessly manage data flow between the host-side LSM and device-side key-value interface, ensuring consistency and high performance.

\item Our approach demonstrates that by introducing an additional storage interface, separate from the traditional block interface, on a singular storage device in the system, we can provide an architecturally beneficial solution to address the inherent limitations of LSM-KVS, which required intentionally lowering the quality of write service to mitigate write stalls.
\squishend

Our extensive evaluation using $db\_bench$~\cite{dbbench} demonstrates that \kcache{} completely eliminates write halts and achieves up to a 17\% increase in throughput compared to the state-of-the-art solution by utilizing underutilized PCIe bandwidth during write stall periods, all while maintaining read performance.
These results show that \kcache{} not only alleviates the performance bottlenecks of existing LSM-KVS systems but also introduces a novel, architecturally superior, cost-effective solution for optimizing write-intensive workloads in modern storage environments.
\section{Background and Related Work}
\label{sec:back}

This section reviews the compaction process in LSM-tree structures, where write stalls occur, and examines related research aimed at mitigating these stalls.

\subsection{Log-Structured Merge Tree and Write Stall Issue}

\begin{figure}[!t]
\centering
\includegraphics[width=1\linewidth]{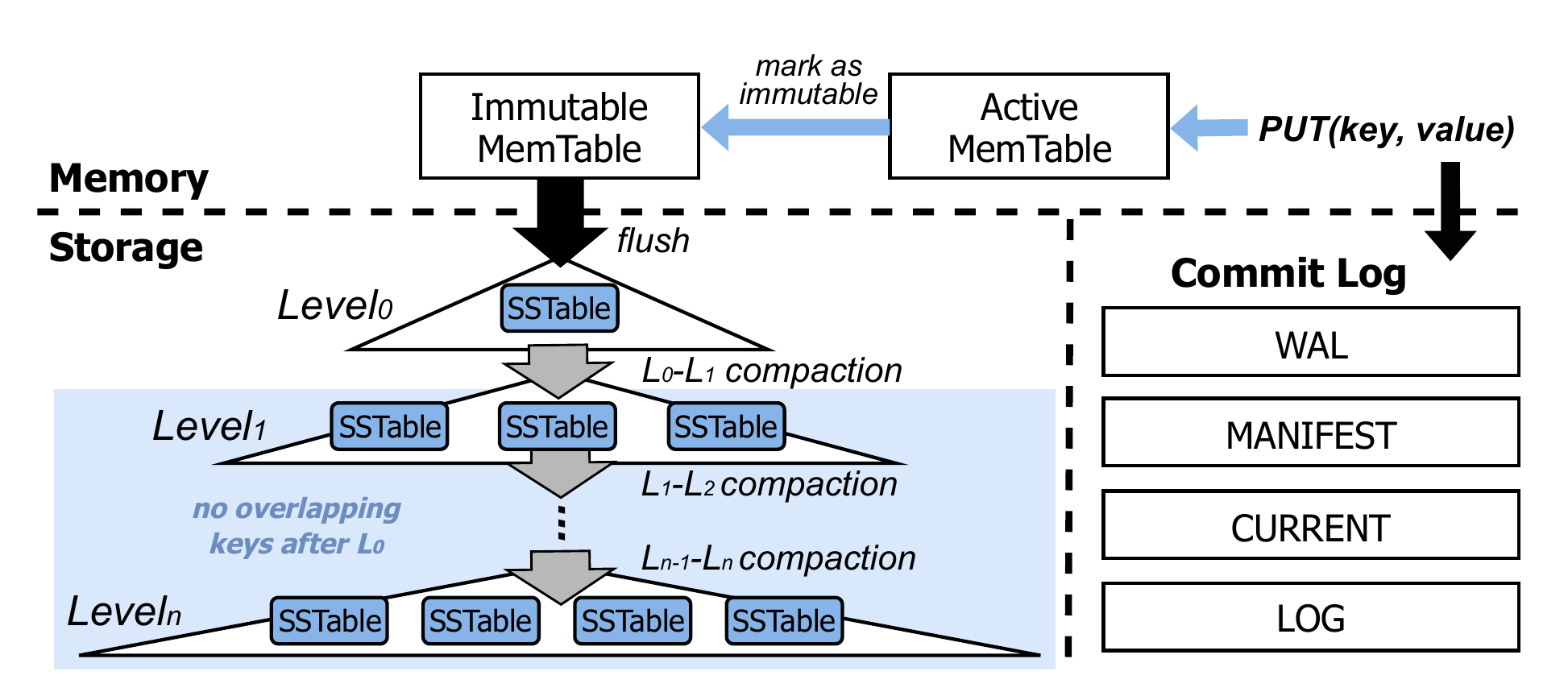}
    \vspace{-15pt}
    \caption{An architecture of LSM-tree.
	}
    \vspace{-8pt}
	\label{fig:back_lsm}
\end{figure}

{The Log-Structured Merge (LSM) tree~\cite{o1996log} is a write-optimized data structure widely adopted in various NoSQL databases including LevelDB~\cite{levelDB}, RocksDB~\cite{rocksdb}, and Cassandra~\cite{cassandra}. The LSM-tree organizes data into memory and disk-based components with hierarchical levels increasing in size, as shown in Figure~\ref{fig:back_lsm}. The memory components include active MemTable (MT) which absorbs the incoming write requests from application. Once MT reaches a size threshold, a new active MT is allocated and old MT is converted into an Immutable MemTable (IMT). The flush operation picks the IMT and convert it into Sorted String Table (SST) file and write to storage device. The SSTs are organized in ascending levels, with each level having a size threshold. When a level reaches the size threshold, SSTs of current victim level $n$ goes through a merge-sort operation, known as compaction, with SSTs of level $n+1$. 
This process ensures key-value pairs within each SST to be sorted and unique.}

\textbf{Write Stall Problem:}
{Despite being write-optimized, LSM-KVSs suffer from the write stall problem. We define the write stall problem as blocking of incoming write requests by the internals of LSM-tree. SILK~\cite{silk} and ADOC~\cite{yu2023adoc} categorize these write stalls into three different events. \bcircled{1} Flush-based write stalls: when the flush operation is not able to keep up with the rate of incoming write requests resulting in exhaustion of memory. \bcircled{2} \llc{}-based write stalls: the SSTs in level $0$ can hold overlapping key range, which necessitate the compaction operation to be serialized between \llc{}. This serialization of \llc{} can lead to blocking of flush operation when L\textsubscript{0} reaches its size threshold, resulting in \llc{}-based write stall event. \bcircled{3} Pending compaction bytes-based write stall: when the lower levels of LSM-KVS delays the compaction operation leading to high space amplification, resulting in blocking of incoming write requests. 
}

\subsection{Existing Optimizations for Addressing Write Stall Issue}

To optimize LSM-KVS, there have been extensive research conducted by academia and industry which can be classified into two categories: (i) software-level and (ii) hardware-level.

\textbf{Software-Level Optimization: }
SILK~\cite{silk} introduces an I/O scheduler that mitigates write stalls by delaying flush and compaction operations to low-load periods, prioritizing flushes and lower-level compactions, and preempting compactions. Despite these strategies, SILK offers minimal performance improvement and exhibits ordinary tail latency under sustained write-intensive and long peak workloads.
RocksDB~\cite{rocksdb} indeed employs a slowdown mechanism~\cite{rocksdb_slowdown} that predicts potential write stalls and intentionally lowers the write throughput to prevent sudden performance drops, but this comes at the cost of increased latency and degraded service quality during heavy workloads.
Blsm~\cite{sears2012blsm} proposes a merge scheduler to coordinate compactions across multiple levels, but the \llc{} still severely stalls foreground requests.
The state-of-the-art solution, ADOC~\cite{yu2023adoc}, also reduces and restores the write ratio as needed, and introduces a new mechanism to dynamically adjust write buffer size and background threads during write-intensive workloads, demonstrating superior write stall mitigation compared to existing solutions.

Despite these efforts, \textit{existing approaches aim to minimize compaction time to mitigate the write stall issue, but they ultimately rely on intentionally lowering the write request rate.}
This trade-off negatively impacts service quality for users, highlighting the inherent limitation of ensuring uninterrupted write operations at the cost of degraded performance.

\textbf{Hardware-Level Optimization: }
To eliminate the high storage stack overhead during key-value writes and compaction, some studies have implemented LSM-KVS directly on SSDs, referred to as Key-Value SSDs (KV-SSDs)~\cite{KVSSDsamsung, jin2017kaml, lee2019ilsm, im2020pink, iteratorkvssd, bandslim}. 
iLSM~\cite{lee2019ilsm} bypasses the file system and block layer within the kernel, thereby improving the I/O latency and throughput of key-value clients. 
PinK~\cite{im2020pink} proposed a resource-efficient LSM-KVS within the KV-SSD and demonstrated that KV-SSDs can reduce CPU and DRAM resource usage on both the host and device side.
In contrast, there are studies that optimize LSM-KVS by leveraging Persistent Memory (PM). 
MatrixKV~\cite{yao2020matrixkv} observes the shortcoming of the original SST format and points to slow \llc{} as the root cause of write stalls when deploying PMs. 
It redesigns the format of SST for PM and proposes a new compaction scheme between the first two levels, which they call column-compaction.
Zhang et. al.~\cite{fpga_compaction} proposed an FPGA-based acceleration engine for LMS-KVS to speed up the compaction process via hardware-software collaboration, improving throughput and averting resource contention.
However, these studies face significant limitations in terms of applicability, as they rely on new devices that \textit{either require completely bypassing host-side LSM-KVS stacks or adding new hardware components}.

\begin{figure}[!t]
    \centering
    \includegraphics[width=0.92\linewidth]{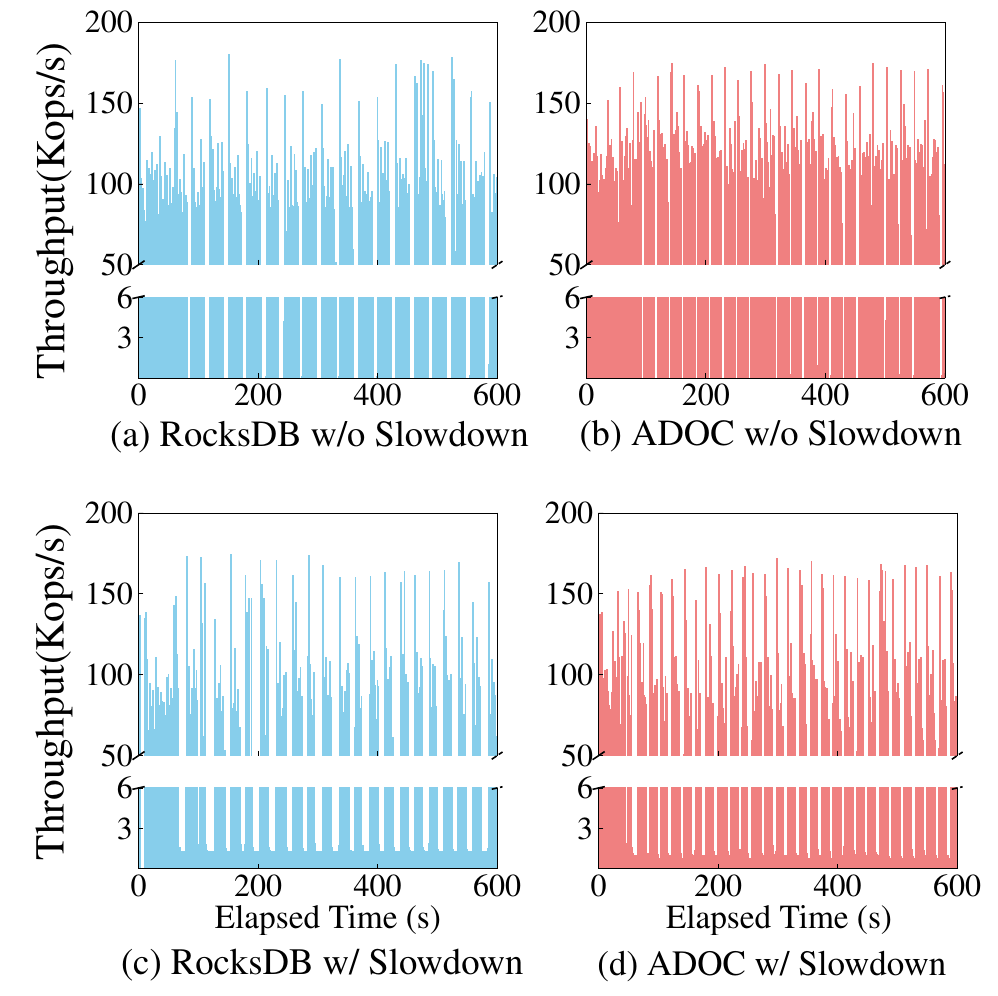}
    \vspace{-4pt}
    \caption{Per-second throughput time-series for RocksDB and ADOC, based on write slowdown usage.}
    \label{fig:motiv_comp_slowdown}
    \vspace{-4pt}
\end{figure}

\section{Problem Definition}
\label{sec:motiv}

In this section, we point out that both industry standard and state-of-the-art software-level solutions both rely on the write slowdown, an inefficient write stall prevention method. 
Furthermore, we highlight that during write stalls in LSM-KVS, the storage device is underutilized, even though it still has the capacity to process I/O requests.

\begin{figure}[!t]
    \centering
    \includegraphics[width=0.92\linewidth]{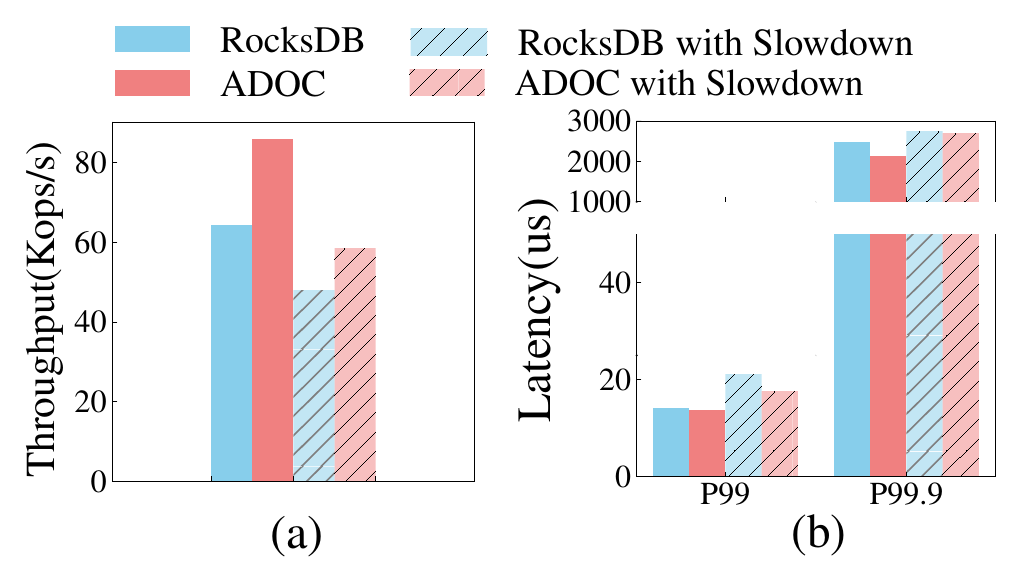}
    \vspace{-4pt}
    \caption{Throughput (a) and tail latency (b) results of RocksDB and ADOC, based on write slowdown usage.}
    \label{fig:motiv_comp_result_slowdown}
    \vspace{-4pt}
\end{figure}

\subsection{Slowdowns: The Inefficient Write Stall Solution}
To prevent write stalls, the basic and most primitive solution is to slow down the writes itself before a write stall occurs. This is done by putting the write thread to sleep for a short duration of time, such as 1~ms\cite{writestall}. Industry standard LSM-KVS such as RocksDB make liberal use of slowdowns during heavy write workloads to prevent write stalls. Meanwhile, ADOC~\cite{yu2023adoc}, the state of the art solution, still falls back to slowdowns as a last resort despite software optimizations such as dynamic allocation of compaction threads and batch size.

To measure the effectiveness and frequency of write slowdowns, we used RocksDB's benchmark tool, $db\_bench$~\cite{dbbench}, and executed \texttt{fillrandom} workload for 600 seconds on RocksDB and ADOC. The experiments were conducted using an OpenSSD-based SSD prototype mounted with a traditional block based interface with the ext4 file-system. The SSD supports a peak bandwidth of approximately 630 MB/s, and is connected to the host via a PCIe Gen2.0 x8 interface, yielding a theoretical maximum PCIe bandwidth of 4 GB/s. Details about our experimental environment are provided in Section~\ref{sec:expr_setup}.

Figure~\ref{fig:motiv_comp_slowdown} (a)-(d) show RocksDB and ADOC's time-series throughput. Here, we present two variants of RocksDB and ADOC, first where the slowdown feature of RocksDB and ADOC is disabled while in the second, the slowdown feature is enabled, respectively. 
Comparing Figures~\ref{fig:motiv_comp_slowdown} (a) and (c) with (b) and (d), respectively, we observe that when the slowdown feature is enabled for both RocksDB and ADOC, the issue of write throughput dropping to zero—i.e., write service halting momentarily—disappears. 
Instead, although the throughput is slightly lower, it remains stable at a base level, providing consistent service at up to 2~Kops/s.
This demonstrates that the slowdown feature effectively mitigates write stalls, ensuring stable and uninterrupted service.
However, it also highlights that the extent of throughput mitigation achieved through the slowdown mechanism is not particularly significant.

Surprisingly, looking at Figure~\ref{fig:motiv_comp_result_slowdown} reveals that these slowdowns actively harm performance in both throughput and P99 latency. While slowdown is in effect, the overall throughput of RocksDB and ADOC dropped by 34\% and 47\% respectively. Tail latency values were also elongated by 48\% and 28\% for RocksDB and ADOC respectively as well. Taking a more microscopic look into the slowdowns, we find that during the workload execution, RocksDB and ADOC experienced a total of 258 and 433 instances of write slowdowns, respectively.
Additionally, ADOC also makes use of more CPU resources over RocksDB while still suffering write slowdowns, as seen in Figure \ref{fig:eval_avg_throughput} in the evaluation section.
As slowdowns ultimately throttle write operations over the course of the workload, the performance results inevitably suffer in comparison to results that do not employ slowdown. In addition, the hit in latency performance can be traced to each slowdown causing the write thread to sleep for a short period, worsening write response times.
In other words, while the slowdown mechanism alleviates the occurrence of write stalls, it ultimately degrades the overall write performance of LSM-KVS, causing users to experience this performance drop.

\begin{figure}[!t]
\centering
\includegraphics[width=1.0\linewidth]
{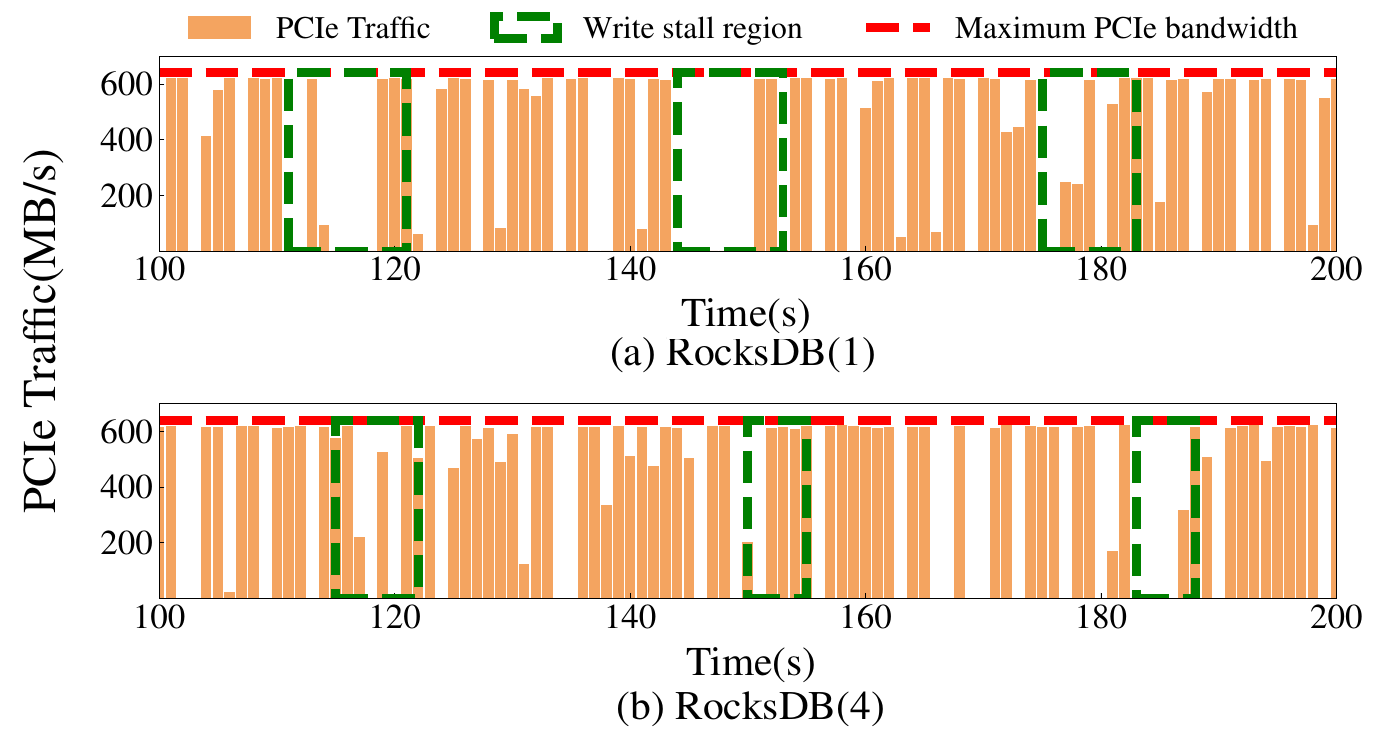}
    \vspace{-13pt}
    \caption{Measurements of PCIe bandwidth utilization in the 100–200 seconds of execution range in RocksDB without applying slowdown techniques.}
    \label{fig:motiv_traffic_elapsetime}
\end{figure}

\subsection{Underutilized PCIe Bandwidth and Device Resources}
In LSM-KVS during a write stall, all user write operations are blocked to allocate system resources for the compaction process. Once compaction is initiated, SSTables (SSTs) are loaded from the storage device/SSD to memory, where a merge-sort operation is performed. Newly created SSTs are then written back to the storage device. 
Importantly, during the merge-sort phase, no data transfer occurs between host's memory and the storage device. This leaves an interval of time during a write stall where potential transfer bandwidth is being unused, yet write operations are not proceeding. 

To empirically verify this behavior, the used PCIe bandwidth of the previous \texttt{fillrandom} experiments were measured while measuring PCIe bandwidth at 1-second intervals using Intel PCM~\cite{intelpcm}. Note that since ADOC's work depends on write slowdowns for its performance optimizations, they were excluded from these experiments.
Figure~\ref{fig:motiv_traffic_elapsetime} illustrates the time-series PCIe bandwidth measurements for RocksDB without slowdown, focusing on a 100-second segment of the total experiment duration. 

Figure~\ref{fig:motiv_traffic_elapsetime} (a)-(b) show the results when using one compaction thread (RocksDB(1)) and four compaction threads (RocksDB(4)), respectively.
The red dotted lines in the figure indicate the maximum bandwidth of the SSD (630~MB/s), while the green dotted boxes mark periods of write stalls. 
From the figure, significant unused bandwidth can be observed during the write stall periods from both configurations of RocksDB.

\begin{figure}[!t]
\centering
\includegraphics[width=0.98\linewidth]
{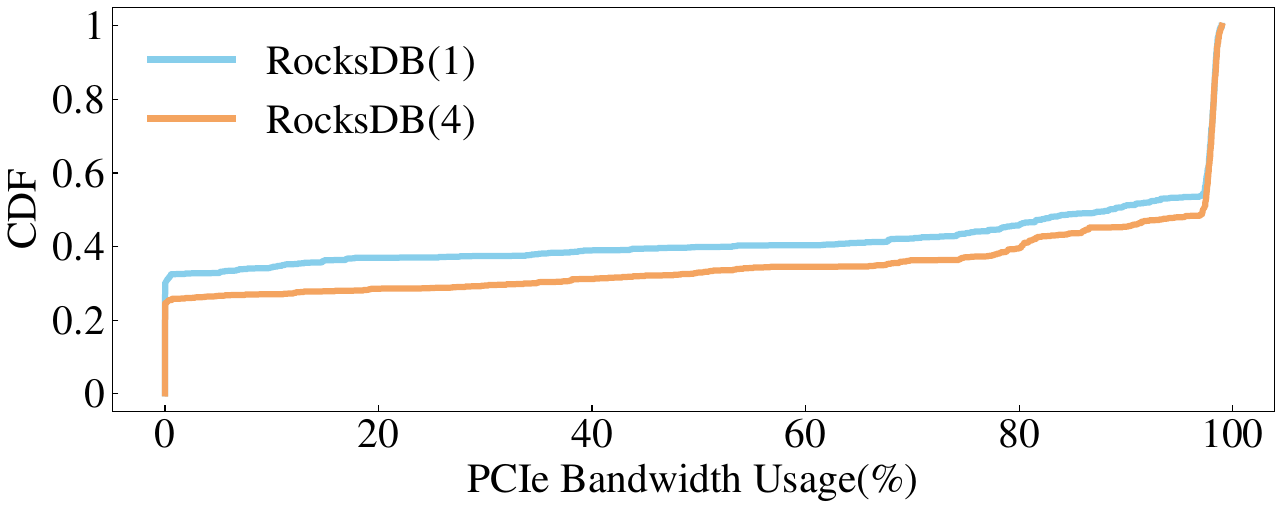}
    \vspace{-4pt}
    \caption{A Cumulative Distribution Function (CDF) of PCIe bandwidth during a period of write stall on RocksDB. The numbers on the legend denote compaction thread count.} 
    \label{fig:motiv_pcie_cdf}
    \vspace{-4pt}
\end{figure}

To further analyze the above, we conducted a statistical analysis of the PCIe bandwidth observed during the write stall periods over the entire 600-second experiment. Figure~\ref{fig:motiv_pcie_cdf} presents the cumulative distribution function (CDF) of PCIe bandwidth utilization during these periods. 
In RocksDB, with one compaction thread, 30\% of the write stall periods exhibit no PCIe bandwidth usage, while 49\% use over 90\% of available PCIe bandwidth. Four compaction threads improve usage somewhat, where 21\% of the write stall periods exhibit no PCIe bandwidth usage and 55\% use over 90\% of available PCIe bandwidth. While one compaction thread does leave more periods of completely no PCIe bandwidth usage during a write stall, both configurations leave up to 90\% of available PCIe bandwidth around 50\% of the time during a write stall. Therefore, these results demonstrate that RocksDB in both configurations leaves a significant portion of the device’s available PCIe bandwidth underutilized during write stalls.

\subsection{Exploring Available I/O Processing Capacity of the SSD}
From the previous experimental results, the following observations can be made.

\begin{tcolorbox}
\squishlist
    \item[\textbf{Observation 1.}] Both state of the art and industry standard solutions make use of write slowdowns to prevent write stalls, which cause a sharp drop in overall throughput and tail latency.
    \item[\textbf{Observation 2.}] PCIe bandwidth is under-utilized during write stalls in industry standard LSM-KVS due to the compaction operation blocking device I/O.
\squishend
\end{tcolorbox}

These observations lead to a dilemma between two currently possible paths. One can choose to keep slowdowns on and maintain I/O operation service, while coming at the great cost of throttling throughput and deteriorating tail latency. On the other hand, one can disable slowdowns and run the LSM-KVS at maximum capacity, keeping throughput and tail latency alive. However, this leads to enlonged write stalls to occur unpredictably.

The experiments also show a third potential path with discovery of underutilized PCIe and device bandwidth during write stalls. This under-utilization is due to the key-value store halting I/O operations while compaction is in progress. If this underutilized bandwidth can be leveraged in times when the SSD still has available I/O processing capacity, the potential to mitigate write stalls and increase performance without sacrificing system resources can be realized.

\begin{figure}[!t]
\centering
\includegraphics[width=1\linewidth]{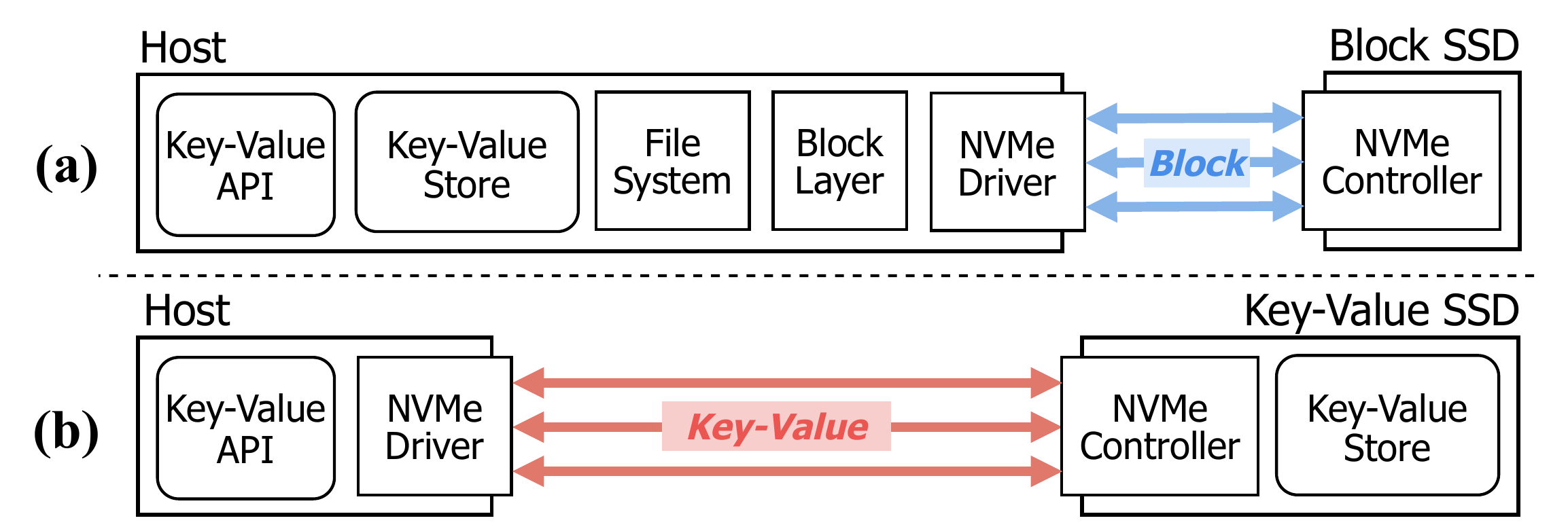}
    \vspace{-15pt}
    \caption{A comparison of software stacks for (a) NVMe Block Interface SSD and (b) NVMe Key-Value Inteface SSD.}
    \vspace{-8pt}
	\label{fig:back_kvssd}
\end{figure}

\section{Opportunities in Key-Value Interfaces}
\label{sec:motiv_kvssd}

Currently, KV-SSDs leverage NVMe extensions~\cite{nvme_kv_spec} to support its key-value interface.
The NVMe-based key-value interface API typically supports point and range queries~\cite{iteratorkvssd}, such as PUT, GET, SEEK, and NEXT, and additionally offers buffered I/O capabilities like compound commands~\cite{compoundcommand}.
As described in Figure~\ref{fig:back_kvssd}, the key-value interface enables efficient I/O processing by eliminating the need for file systems and block layers, effectively simplifying the storage software stack and reducing the overhead associated with multi-layer space management during processing writes and compaction.

The KV-SSDs share the same NAND flash address space and use the same Flash Translation Layer (FTL) mechanisms as traditional block-based NVMe SSDs but internally implement a LSM-KVS at the controller level~\cite{bandslim,lee2019ilsm,im2020pink}. 
The controller abstracts logical addresses for point and range query executions, enabling direct key-value service within the device. 
Aside from executing point and range queries internally, \textit{the rest of the storage infrastructure, such as the NVMe interface and FTL-managed logical-to-physical address mapping, remains identical to that of conventional SSDs}, ensuring compatibility while offering enhanced functionality. 
Based on this, we propose designing a hybrid dual-interface SSD that supports both block and key-value interfaces. 
This approach allows the SSD to leverage the available bandwidth and processing capacity during write stalls in LSM-KVS systems. 
By temporarily redirecting pending write requests through the key-value interface, the SSD can reduce the impact of write stalls and improve overall performance without disrupting ongoing operations in the LSM-KVS.
\section{Design of \kcache{}}
\label{sec:design}

This section introduces the design objectives of \kcache{}, details the hardware and software components involved in its implementation, explains their operation, and discusses how crash consistency is ensured.

\subsection{Design Goals}

To address the aforementioned issues, we propose \kcache{}, a novel hybrid hardware-software co-design framework that leverages a dual-interface SSD architecture to eliminate write stalls and optimize the utilization of storage bandwidth.  
The design goals of \kcache{} are as follows:

\noindent\textbf{\textit{G1}. Mitigating Write Stalls Effectively:} Leverage the key-value interface of the hybrid SSD to serve as a temporary in-device write buffer during host-side write stalls. 
By redirecting writes to the key-value interface, \kcache{} can prevent the host-side LSM from becoming overloaded during compaction.

\noindent\textbf{\textit{G2}. Maximizing I/O Bandwidth Utilization:} Ensure that the SSD’s available bandwidth and I/O processing capacity are fully utilized during write stalls by dynamically switching between the block and key-value interfaces. 

\noindent\textbf{\textit{G3}. Seamless Integration for Consistency and Performance:} Achieve seamless integration between the hybrid SSD and host LSM-KVS by employing efficient metadata management and a rollback mechanism. 
This ensures data consistency between the host’s LSM and the device’s key-value write buffer, even when switching between interfaces. 

\begin{figure}[!t]
	\centering
    \includegraphics[width=1.00\linewidth]{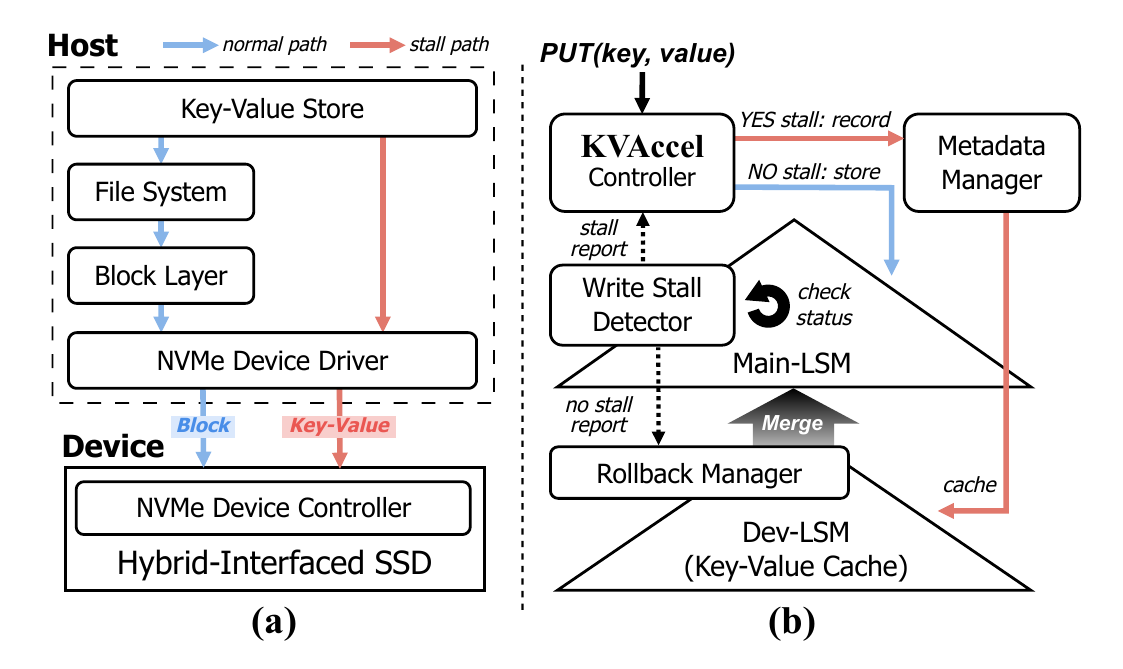}
    \vspace{-18pt}
    \caption{(a) A software stack of \kcache{} and (b) The write path of \kcache{} shown using its software modules.} 
    \vspace{-8pt}
	\label{fig:design_overview}
\end{figure}

\subsection{Overall Architecture}

\noindent\textbf{Hardware and Software.}
\kcache{} is system that offers dynamic redirection and rollback techniques to a LSM-KVS to both mitigate write stalls and fully utilize available I/O bandwidth. This is achieved through the close co-design of software and hardware components. 
The \textit{Software} components assign I/O commands to the correct interface depending on real-time information of the database. Maintaining the consistency of the database between the two interfaces during database operations is also paramount in the software design.
The \textit{Hardware} components implements the disaggregation of separate block and key-value interfaces to allow for the hybrid interface of the SSD. The hardware also implements support for bulk range scan operations over its write buffer to perform the rollback operation for consistency of our system.

\noindent\textbf{Disaggregation and Aggregation.}
The design of \kcache{} is based on two key factors: disaggregation and aggregation. 
\textit{Disaggregation} facilitates the division of the SSD into the hybrid interface, as well as the software required for the I/O pathways for each interface. 
\kcache{} disaggregates a SSD into a hybrid interface with separate block and key-value interfaces, each with its own separate LSM-tree that each interface manages. 
\textit{Aggregation} focuses on managing the data stored in the hybrid interface SSD as if it were a single database instance.
This includes unifying the host-side and device-side LSM-trees during rollback operations by efficiently merging cached key-value pairs from the device back into the host’s LSM structure. 
Additionally, \kcache{} maintains a global metadata manager to track the locations of key-value pairs across both interfaces, ensuring transparent access to data regardless of its physical placement in the SSD.

Figure \ref{fig:design_overview}(a) shows the potential of writing using the key-value interface during periods of write stall. Through the key-value interface, I/O operations can bypass the file system and block layer and drill a path straight to the NVMe controller via the driver. This path offers a method to service I/O requests uninterrupted through the key-value interface, even while write stalls are occurring on the database running on the block interface.
A key point in disaggregation is that the hybrid interface of both block and key-value interfaces are implemented in a singular device. This is significant in that to see the benefits of \kcache{}, only the one storage device programmed to run is required. 
A single device solution enables \kcache{} to bypass the burdens of additional hardware deployment (e.g., PM, FPGA) that previous hardware-level solutions to the write stall issue introduce.

\subsection{Interface Pathing via Software Modules} \label{sec:modules}

To make use of the hybrid interface, the decision to use which interface needs to be made every time a operation is requested by the database. 
To do this, \kcache{} makes use of the following four software components shown in Figure \ref{fig:design_overview}(b) to make this decision to make full use of unused device bandwidth.
The LSM-tree residing on the block interface is labeled \textit{Main-LSM}, while the LSM-tree on the key-value interface labeled \textit{Dev-LSM}. 
Main-LSM is used by the LSM-KVS running on the host machine, and uses the block interface to serve write operations during periods when write stall is not present. 
On the other hand, Dev-LSM runs entirely within the hybrid SSD, and uses the key-value interface to serve write operations when Main-LSM is facing a write stall as secondary cache storage.

\squishlist
\item \textbf{Detector:} 
The Detector periodically checks three components of Main-LSM that are associated with the characteristics of a write stall: the number of SSTs in \lz{}, MT size, and pending compaction size. The Detector then reports this information to the Controller to use for path determination.

\item \textbf{Controller:}
The Controller uses the information reported by the Detector to issue I/O operations to the correct interface. If the Detector reports that no write stalls are occurring, the Controller directs the operation to Main-LSM. If the Detector reports a write stall, the Controller performs the operation to the Dev-LSM.

\item \textbf{Metadata Manager:}
As the SSD has been disaggregated into a hybrid interface, the data written can be in either Main-LSM or Dev-LSM. To keep track of which interface the database needs to use for future read operations, the key-value pairs that are redirected to the Dev-LSM are kept track of. This metadata of a key-value pair's location is captured in a hash table in memory, and is used for membership testing for future operations that need to know the location of a certain key-value pair. In the case of a system failure and data loss of the metadata manager were to happen, the data can be recovered by a range scan covering every key-value pair in the key-value interface.

\item \textbf{Rollback Manager:} 
To aggregate the two LSM-trees into one, returning the cachced key-value pairs from Dev-LSM to Main-LSM is required. To facilitate this, the Rollback Manager is tasked to initiate the rollback operation depending on the contention status of Main-LSM. The Rollback Manager receives information of the presence of a write stall from the Detector. Further details on the rollback mechanism can be found in Section~\ref{sec:rollback}.
\squishend

With these modules, the read and write paths of \kcache{}, depending on the status of the Metadata Manager and the presence of a write stall, can be seen as follows. 

\squishlist 
\item 
\textbf{Read Path:} \textit{(1)} The Metadata Manager checks the location of the queried key. \textit{(2)} If the key-value pair is in the Main-LSM or if the Dev-LSM is empty, the Controller directs the read operation to the Main-LSM. \textit{(3)} If the key-value pair is found in the Dev-LSM, the Controller redirects the read operation to the Dev-LSM. 
\item 
\textbf{Write Path:} \textit{(1)} The Detector checks for the presence of a write stall. \textit{(2)} If a write stall is detected, the Controller, through the Metadata Manager, updates the record to indicate that the key-value pair is now in the Dev-LSM, and the pair is written to the Dev-LSM. \textit{(3)} If no write stall is detected, the Controller directs the key-value pair to be written to the Main-LSM. \textit{(3-1)} If the Metadata Manager indicates that an overlapping key-value pair already exists in the Dev-LSM, it updates the record to indicate that the latest key-value pair is now in the Main-LSM. 
\squishend

Note that these paths only refer to the point queries of \texttt{Put()} and \texttt{Get()}. For range queries, refer to Section~\ref{sec:range_query}. 

\begin{figure}[!t]
	\centering
    \includegraphics[width=1\linewidth]{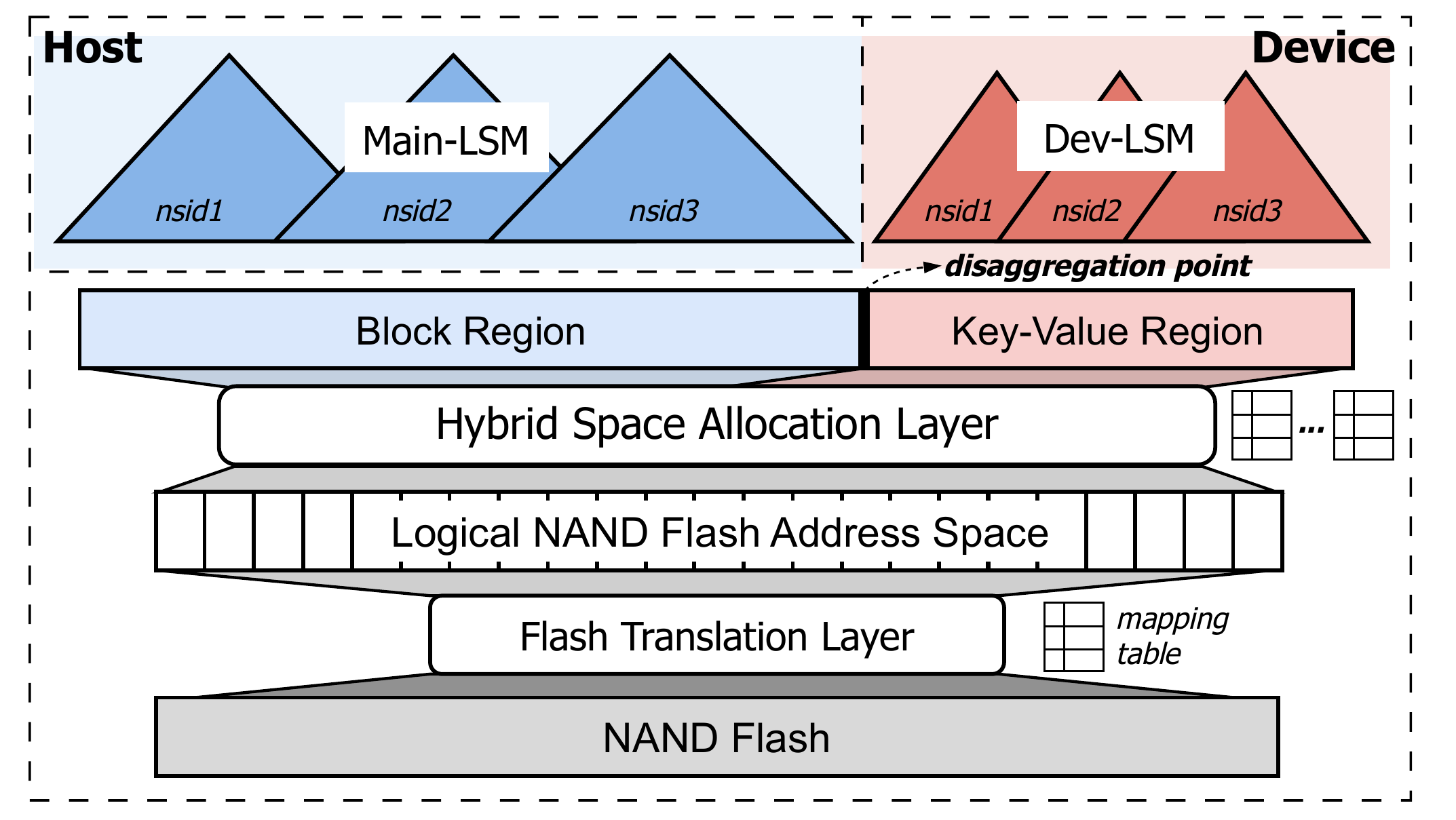}
    \vspace{-17pt}
    \caption{A dynamic, namespace-aware hybrid NAND flash space allocation of disaggregated NAND flash address space.}
    \vspace{-8pt}
	\label{fig:design_disg}
\end{figure}

\subsection{Hybrid Dual-Interface SSD}

To support a storage device with a dual-interface, the SSD's logical NAND flash address space range is disaggregated into two address ranges, as shown in Figure \ref{fig:design_disg}. One address range is used for the block interface, and the other for the key-value interface. The address ranges are defined by the disaggregation point, which is a logical address that defines the end of one interface and the start of the next. The SSD's controller issues different commands for each respective interfaced based on the given opcode of the NVMe command. Block interface commands perform FTL mapping over the logical address space allocated for the block interface. Key-value interface commands allocates NAND pages from the logical address space allocated for the key-value interface. Both interfaces make use of existing NVMe command set specifications made for each respective interface type~\cite{nvme_spec, nvme_kv_spec}.

As the FTL maps logical address spaces for each interface separately, there are no issues of overlapping logical NAND pages between the two interfaces. When a file system is created for the block interface, the file system only sees the address range that was allocated for the block interface, and reports storage capacity to reflect said address range. Likewise, the key-value interface will only store key-value pairs up to the limits of its allocated address range.

\textbf{Multi-Tenancy Support:}
The ability to create multiple isolated regions on a singular device is a key requirement in multi-tenancy environments. To offer multi-tenancy in \kcache{}, both the block and key-value interface needs to support such features of isolated divisions. Multi-tenancy on the block interface is supported by namespaces as specified in the NVMe standard \cite{nvme_namespaces}, while previous works on supporting namespaces on the key-value interface \cite{isokvssd} are compatible with \kcache{}'s key-value interface implementation. 
By utilizing both namespace implementations for each interface and matching namespaces in both interfaces for each tenant, \kcache{} can fully support the requirements of multi-tenancy with both interfaces.

\subsection{Rollback Operation} \label{sec:rollback}

\begin{figure}[!t]
	\centering
    \includegraphics[width=1\linewidth]{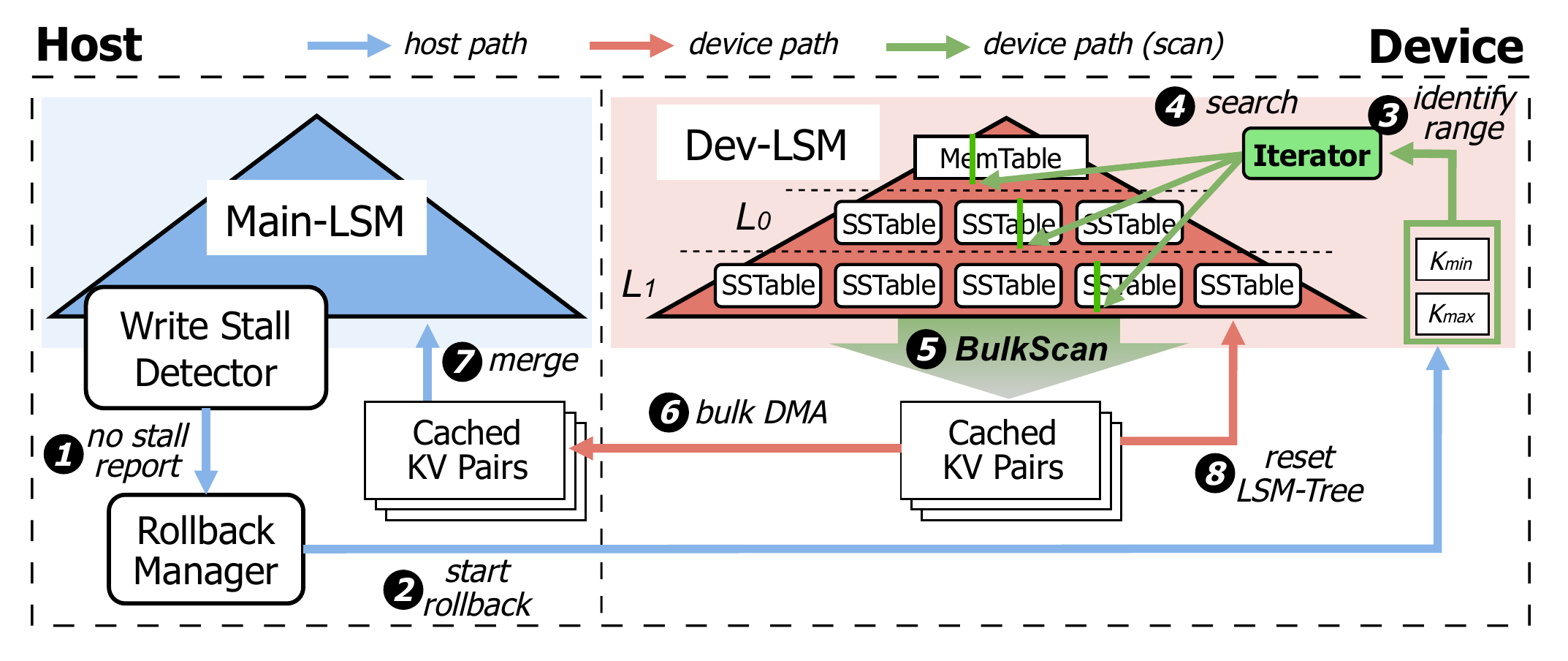}
    \vspace{-17pt}
    \caption{An in-device iterator-based range scan to accelerate host-device co-managed rollback mechanism in \kcache{}.} 
    \vspace{-10pt}
	\label{fig:design_rollback}
\end{figure}

To return the two separated LSM-KVSs back into a singular database, the cached key-value pairs in Dev-LSM needs to be returned back to Main-LSM. This is done in a process called rollback. Figure \ref{fig:design_rollback} displays an overview of the rollback operation, and the interactions between the host and the device during said operation.
A rollback operation starts with the Detector and Rollback Manager. \bcircled{1} As rollback is only performed during periods when write stall is not present in Main-LSM, the Detector needs to notify the Rollback Manager the appropriate moment to start rollback. 
\bcircled{2} When no write stalls are detected and there are key-value pairs in Dev-LSM, the rollback operation is initiated.

\textbf{Rollback Scheduling:} 
The rollback manager can schedule a rollback \textit{eagerly} or \textit{lazily} depending on the characteristics of a workload. 
An eager rollback scheme will trigger rollback as soon as the rollback manager detects that there are enough leftover resources in the LSM-KVS. 
Such a scheme is better suited for read oriented workloads, as point read query on the Dev-LSM are much slower than its counterpart in the Main-LSM, as such a read operation requires querying the slower device storage every time for a read operation.
On the other hand, a lazy rollback scheme will trigger rollback when it is certain no other workload will interfere or be interfered by the rollback. This scheme is designed for write intensive workloads, as there is little penalty to keep the key-value pairs in Dev-LSM in this workload and therefore less urgency to perform rollback.

\textbf{Iterator-Based Bulky Range Scan:} Regardless of the chosen rollback scheduling scheme, rollback needs to be performed as fast as the system allows. This is mainly due to the possibility of I/O operation conflicts. Performance can especially be crippled in cases where read and write operations happen simultaneously, where a time-consuming read operations can impact write operations. Such a conflict can occur with the aforementioned slower point read query on the Dev-LSM. 
To accelerate the rollback operation, \bcircled{3} an iterator first identifies the range of the entire Dev-LSM to perform a range query by using the start and end keys of Dev-LSM. \bcircled{4} The iterator will search over the entire Dev-LSM, and \bcircled{5} cache key-value pairs are serialized in bulk and transferred to host via device memory using NAND I/O.
\bcircled{6} key-value pairs are then saved to system memory in chunks of 512~KB, so that the host can access the key-value pairs using Direct Memory Access (DMA). This size was chosen as 512~KB is the maximum size unit that DMA supports for data transfer on our platform. \bcircled{7} Finally, the host can retrieve and unpacks the key-value pairs to merge back in Main-LSM. 
\bcircled{8} After one rollback operation is finished, a reset is performed on the Dev-LSM to prevent consistency issues in the next rollback operation. By resetting Dev-LSM, the key-value pairs redirected to be involved in the next rollback can be the most up to date data. The reset also ensures the rollback of all key-value pairs in Dev-LSM to be completely written back to Main-LSM.

An important point to keep mind of is as the duration of a write stall is relatively short, Dev-LSM does not have a large amount of SSTs that needs to be rolled back. 
This fact, along with the aforementioned iterator-based range scan method, can ensure that every rollback operation can be finished in between periods of write stall.

\subsection{Range Query Support}
\label{sec:range_query}

\begin{figure}[!t]
	\centering
    \includegraphics[width=1\linewidth]{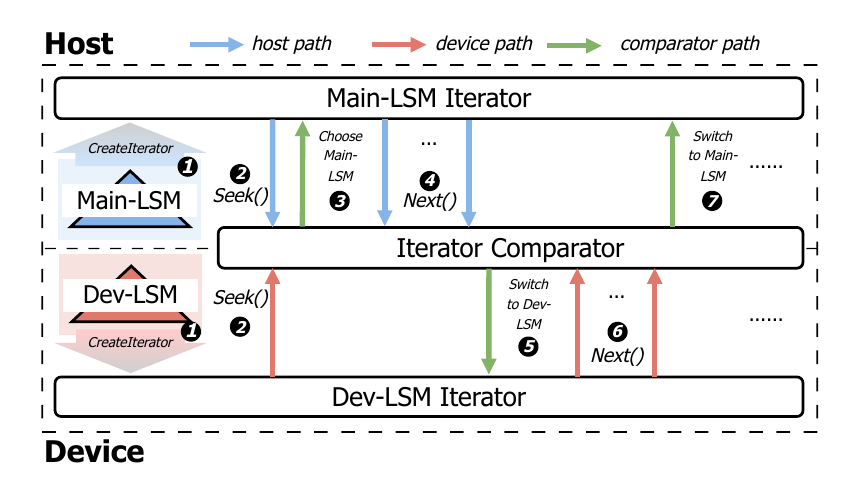}
    \vspace{-17pt}
    \caption{A range query operation in \kcache{}.}
    \vspace{-8pt}
	\label{fig:design_rangequery}
\end{figure}

Range queries work with the combination of iterator implementations of each respective interface in \kcache{}. Main-LSM can use the chosen LSM-KVS's implementation of iterator and range scan. Meanwhile, Dev-LSM's key-value interface has support for iterators and range scan functionality \cite{iteratorkvssd}, and \kcache{} utilizes the same bulky range scan mechanism from the rollback operation. 
Each interface will have its own iterator to perform \texttt{Seek()} and \texttt{Next()} operations over its LSM-trees. 
The two iterators will be aggregated to work in tandem to perform a range query over the entire LSM-KVS. 
An example range query operation is shown in Figure \ref{fig:design_rangequery}. \bcircled{1} An iterator for both Main-LSM and Dev-LSM are created, and \bcircled{2} a \texttt{Seek()} operation is performed for both LSM-trees. \bcircled{3} The values returned from the \texttt{Seek()} operation are sent to the iterator comparator to be compared and saved. The iterator that returned the desired start key, or the smaller key if the desired start key was not found, is selected. \bcircled{4} The selected iterator than procedes to perform \texttt{Next()} operations, until the iterator returns a key larger than the key saved from the opposing iterator's first \texttt{Seek()} operation. \bcircled{5} The used iterator is then switched, and \bcircled{6} \texttt{Next()} operations are continued on the switched iterator. 
\bcircled{7} This process of switching iterators when necessary continues until the desired end point is reached or the final key-value pair is reached. 

\subsection{ACID Property Management}

\kcache{} maintains the ACID properties of database transactions by leveraging its dual-interface SSD design. 
First of all, for atomicity, the disaggregation of NAND flash address space within the dual-interface SSD handles operations between the Main-LSM and Dev-LSM in a completely independent manner. 
The rollback manager then monitors and reverts any changes made during incomplete transactions, ensuring that any partial or failed transactions during write redirection or rollback are consistently cleaned up by a rollback manager. 
Consistency is upheld through real-time metadata tracking and validation across both interfaces, with a dynamic consistency checker enforcing strict rollback protocols during high-pressure situations to maintain data accuracy in Main-LSM. 
The Metadata Manager directs all read and write operations to the appropriate structure, ensuring a seamless transition from Dev-LSM to Main-LSM.
To achieve isolation, \kcache{} segregates concurrent I/O operations between the two LSM structures through the Controller Module, isolating Dev-LSM as a temporary cache during write stalls and preventing interference between the interfaces. 
Each range query is executed independently with separate iterators for each LSM, thereby ensuring query consistency even during ongoing write operations.
Durability is guaranteed through a two-stage commit protocol that writes data first to Dev-LSM’s non-volatile NAND space before committing it to Main-LSM. 
This method secures committed transactions even during unexpected power failures or system crashes. 
In the event of a failure during rollback, the data remains in Dev-LSM until the system is restored, ensuring no loss of committed transactions. 
This robust architecture makes \kcache{} capable of maintaining database integrity and performance under various system conditions.
\section{Evaluation} \label{sec:eval}

\subsection{Experimental Setup}
\label{sec:expr_setup}

We implemented \kcache{}'s hardware components by extending the state-of-the-art NVMe KV-SSD~\cite{lee2023iterator} based on Cosmos+ OpenSSD platform~\cite{cosmos_plus_open_ssd}. The SoC of the platform operates the \kcache{}'s hybrid-interface SSD controller, the PCIe interface controller, the DRAM controller, and the NAND flash controller. A single ARM core of the Cosmos+ is used to run Dev-LSM's I/O operations, as well as other required operations such as flush and compaction operations.
The host system runs a modified version of the Linux kernel to facilitate the hybrid-interface SSD, as well as the NVMe block and key-value interface drivers.
Table~\ref{tab:eval_setup_openssd} and Table~\ref{tab:eval_setup_host} present the hardware and software specifications of our setup.  

\begin{table}[!h] 
    \centering
    \caption{Specifications of the OpenSSD platform.} 
    \vspace{-5pt}
    \label{tab:eval_setup_openssd}
       \footnotesize
       \begin{tabular}{|l||l|}
            \hline
                \cellcolor{gray!15}SoC & Xilinx Zynq-7000 with ARM Cortex-A9 Core 
            \\\hline
            \cellcolor{gray!15}NAND Module & 1TB, 4 Channel \& 8 Way
            \\\hline
            \cellcolor{gray!15}Interconnect & PCIe Gen2 $\times$8 End-Points \\\hline
    \end{tabular}
    \vspace{-10pt}
\end{table}

\begin{table}[!h] 
    \centering
    \caption{Specifications of the host system.}
    \vspace{-5pt}
    \label{tab:eval_setup_host}
       \footnotesize
       \begin{tabular}{|l||l|} 
        \hline
        \cellcolor{gray!15}CPU & \begin{tabular}[c]{@{}l@{}}Intel(R) Xeon(R) Gold 6226R CPU @ 2.90GHz (32~cores),\\CPU usage limited to 8 cores.\end{tabular} \\ 
        \hline
        \cellcolor{gray!15}Memory & 384GB DDR4 \\ 
        \hline
        \cellcolor{gray!15}OS & Ubuntu 22.04.4, Linux Kernel 6.6.31 \\
        \hline
        \end{tabular}
\end{table}

\begin{table}[!t]
\centering
\caption{LSM-KVS configurations. For all figures, the numbers next to each LSM-KVS refer to compaction thread count. For \kcache{}, the settings refer to the Main-LSM.}
\label{tab:lsm-kvs}
\footnotesize
\centering
\begin{tabular}{|c||c|c|} 
\hline
\cellcolor{gray!15}LSM-KVS & \cellcolor{gray!15}Compaction Threads $(n)$ & \cellcolor{gray!15}MT Size \\ 
\hline\hline
\multirow{3}{*}{\kcache{}$(n)$} & 1 & \multirow{9}{*}{128~MB} \\ 
\cline{2-2}
 & 2 &  \\ 
\cline{2-2}
 & 4 &  \\ 
\cline{1-2}
\multirow{3}{*}{RocksDB$(n)$} & 1 &  \\ 
\cline{2-2}
 & 2 &  \\ 
\cline{2-2}
 & 4 &  \\ 
\cline{1-2}
\multirow{3}{*}{ADOC$(n)$} & 1 &  \\ 
\cline{2-2}
 & 2 &  \\ 
\cline{2-2}
 & 4 &  \\
\hline
\end{tabular}
\end{table}

\begin{table}[!t]
\centering
\caption{$db\_bench$ workload configurations. Each benchmark was run with a 4~B key and 4~KB value size. Workload A,B,C were run for 600 seconds, and Workload D performed 60K read operations.}
\label{tab:benchmarks}
\renewcommand{\arraystretch}{1.2}
\resizebox{\columnwidth}{!}{
\begin{tabular}{|c||c|c|c|} 
\hline
\cellcolor{gray!15}Name & \cellcolor{gray!15}Type & \cellcolor{gray!15}Characteristics & \cellcolor{gray!15}Notes (write/read ratio) \\ 
\hline\hline
A & fillrandom & 1 write thread & No write limit \\ 
\hline
B & \multirow{2}{*}{readwhilewriting} & \multirow{2}{*}{\begin{tabular}[c]{@{}c@{}}1 write thread \\+ 1 read thread\end{tabular}} & 9:1 \\ 
\cline{1-1}\cline{4-4}
C &  &  & 8:2 \\ 
\hline
D & seekrandom & \begin{tabular}[c]{@{}c@{}}1 range query thread\\(Seek + 1024 Next)\end{tabular} & \begin{tabular}[c]{@{}c@{}}Run after initial \\20GB fillrandom\end{tabular} \\
\hline
\end{tabular}
}
\vspace{-5pt}
\end{table}

\kcache{}'s software components were implemented on RocksDB v8.3.2. The Detector, Controller, Metadata Manager, and Rollback Manager software modules are all implemented on top of RocksDB. The Detector and Rollback Manager in particular run a thread detached from the RocksDB thread, refreshing the status of Main-LSM and checking for conditions of rollback every 0.1 seconds.  

For performance evaluations, we slightly modified $db\_bench$~\cite{dbbench}, a widely recognized benchmarking tool used in RocksDB. 
We enabled $db\_bench$ to send NVMe key-value commands to the Cosmos+ OpenSSD platform through the NVMe passthrough.
The LSM-KVSs and the configurations used for the evaluations are detailed in  Table \ref{tab:lsm-kvs}. The various patterns of the workloads to verify our proposed design are described in Table \ref{tab:benchmarks}.

\subsection{Write Stall Mitigation Evaluation}
\begin{figure}[!t]
    \centering
    \includegraphics[width=1.0\linewidth]{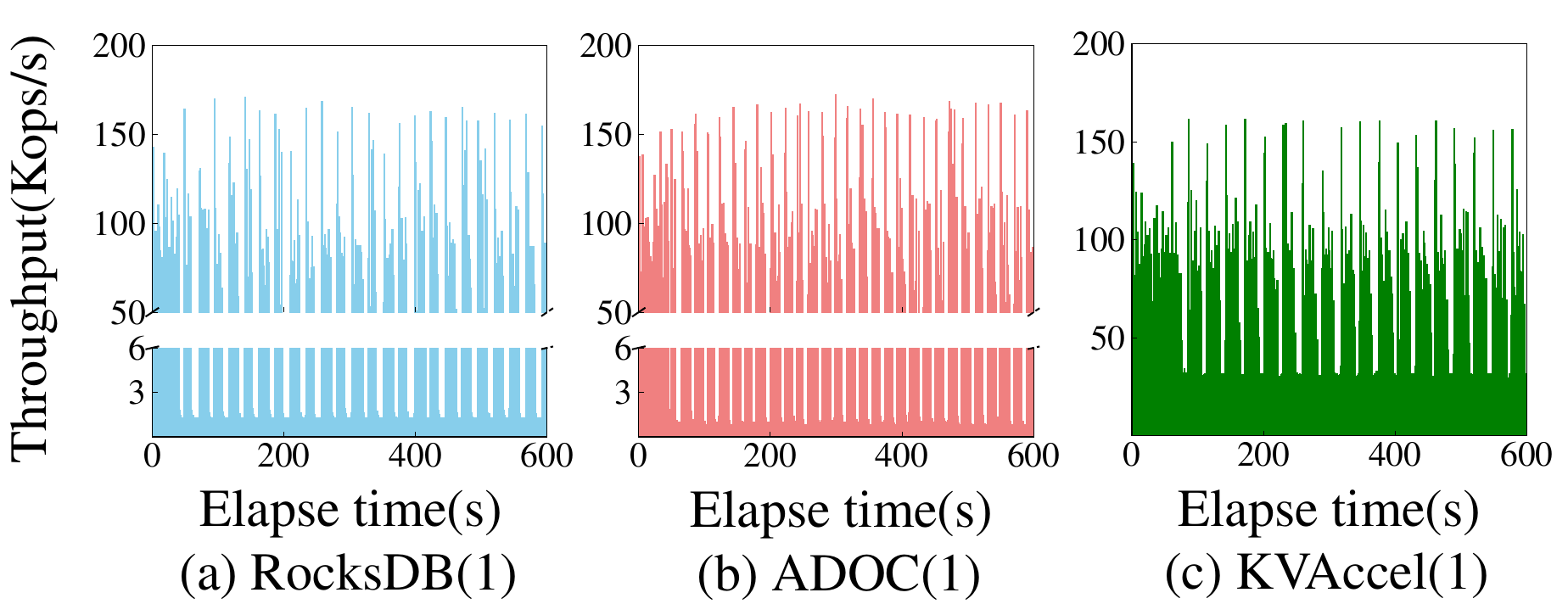}
    \vspace{-16pt}
    \caption{Per-second throughput for each LSM-KVS while running workload A.}
    \vspace{-6pt}
    \label{fig:eval_stall_avoid}
\end{figure}

This section demonstrates \kcache{}'s ability to mitigate write stalls via I/O redirection.
Figure~\ref{fig:eval_stall_avoid} displays the per-second throughput of all three LSM-KVS (RocksDB, ADOC, and \kcache{}) during the entirety of workload A. 
Figure~\ref{fig:eval_stall_avoid} (a) and (b) focus on the periods of lower throughput in order to examine the decrease in throughput that occurred during the slowdown phase.
ADOC and RocksDB can be both seen suffering from slowdowns to 2 Kop/s in order to prevent a write stall. In similar periods, \kcache{} proceeds to write upwards of 30 Kop/s, showing I/O redirection response of \kcache{} allowing for the avoidance of write stalls.

A point to emphasize here is \kcache{} does not employ any slowdown mechanisms to avoid a write stall. This is because \kcache{} is inherently designed to accept writes in its full capacity during a write stall via redirection instead of intentionally throttling write flow to attempt to avoid a write stall. This different approach to the write stall problem allows \kcache{} to maintain write operations while greatly lowering performance compromises, while other LSM-KVSs suffer from slowdowns or face a write stall depending on workload settings.

\subsection{Performance Evaluation}

In this section, the read/write performance, and efficiency of \kcache{} will be demonstrated with the workloads of Table \ref{tab:benchmarks}. Here, we introduce a scoring metric of the ratio between throughput and resources used to better demonstrate the efficiency of throughput in relation to resources used.

\begin{equation}
\text{Efficiency} = \frac{\text{Avg. Throughput(MB/s)}}{\text{Avg. CPU usage(\%)}}
\end{equation}

\begin{figure}[!t]
    \vspace{-4pt}
    \centering
    \includegraphics[width=1\linewidth]{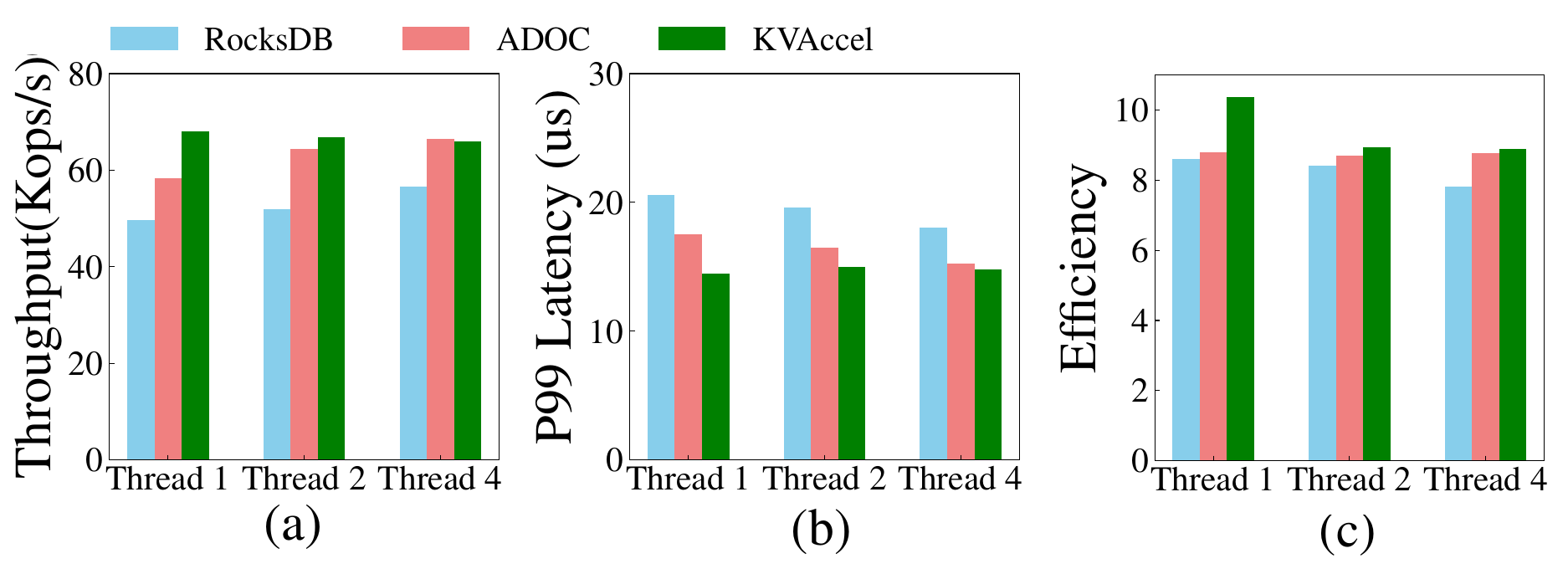}
    \vspace{-16pt}
    \caption{(a) Throughput, (b) P99 Latency, and (c) Efficiency scores of all evaluated LSM-KVS for workload A. Thread counts here denote compaction thread count.}
    \vspace{-6pt}
    \label{fig:eval_avg_throughput}
\end{figure}

Figure~\ref{fig:eval_avg_throughput} shows the average throughput, P99 latency, and efficiency respectively of all LSM-KVS configurations performing workload A. To demonstrate the full potential of \kcache{} in a write-only operation, rollback and compaction operations in Dev-LSM were disabled for workload A. This is because for a write-only workload phase, a lazy rollback scheme that performs rollback after the workload completes is the most sensible option.

\kcache{} shows at most a 37\% and 17\% improvement over its respective configuration in compaction thread count than RocksDB and ADOC, respectively.
In P99 latency, a maximum of 42\% and 20\% decrease in latency was also observed between respective configurations between \kcache{} and RocksDB, ADOC, respectively.
A key point of interest that can be observed is that \kcache{} with only one compaction thread shows similar write throughput with ADOC employing four compaction threads. This is due to the fact that \kcache{} is able to contribute more to throughput when write stalls are longer and happen more frequently. Increasing the compaction thread count ultimately reduces write stall length and frequency, thus lowering the effectiveness of \kcache{}.

Referring to the efficiency metric for the results, \kcache{} also maintains the better efficiencies in host machine's resources between all LSM-KVS compared, with \kcache{}(1) shows the best efficiency over all configurations. This is because \kcache{} is able to achieve the higher throughput results while maintaining the same CPU utilization.

\begin{figure}[!t]
    \centering
    \includegraphics[width=1.0\linewidth]{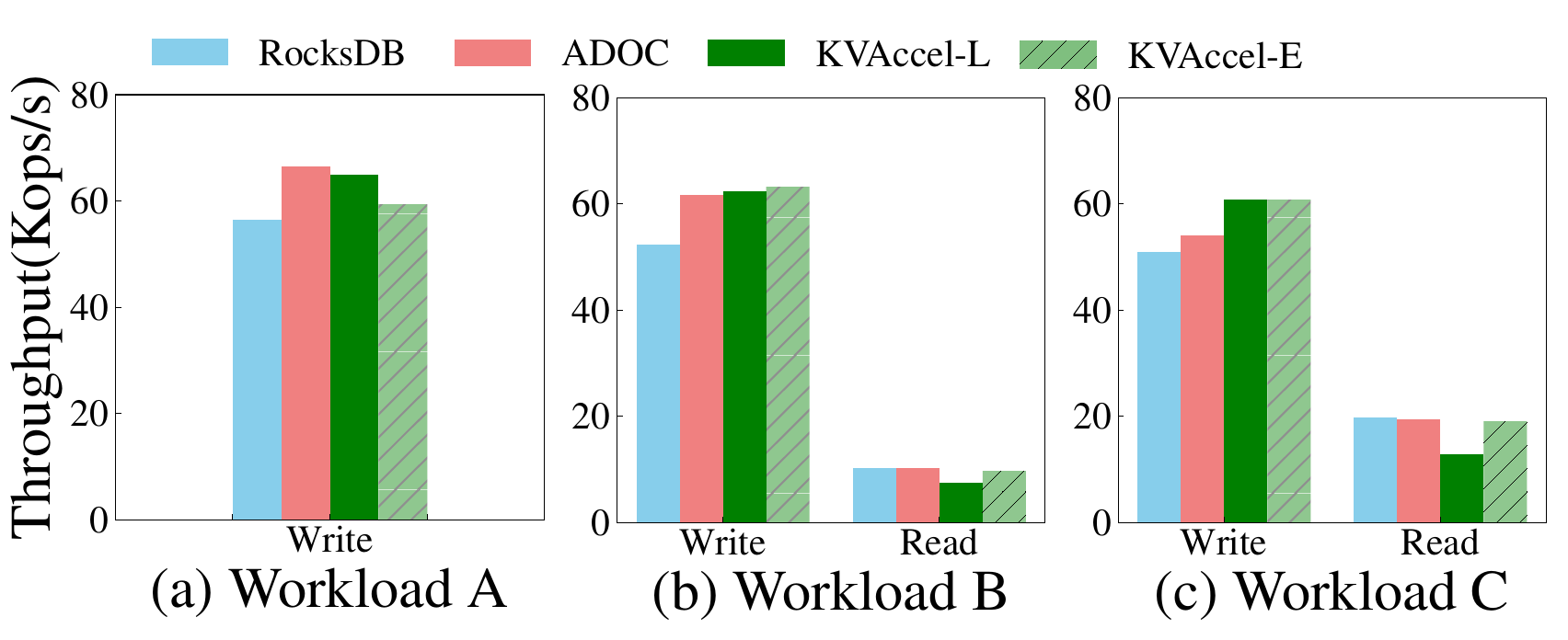}
    \vspace{-16pt}
    \caption{Read and write throughput comparison of different workloads based on rollback schemes choice. \kcache{}-L uses a lazy rollback scheme, and \kcache{}-E uses an eagar rollback scheme.  All LSM-KVS configurations in this figure use 4 compaction threads. Read throughput is non-applicable for workload A, as workload A is a 100\% write workload, and is thus excluded.}
    \label{fig:eval_rww_rollback}
    \vspace{-4pt}
\end{figure}

In order to handle more diverse scenarios, rather than using write- optimized solution, workloads A to C were also performed to evaluate \kcache{} under different rollback schemes. 
The results of these workloads of all LSM-KVS configurations can be seen in Figure~\ref{fig:eval_rww_rollback}, where comparisons of rollback schemes based on workload type are also made. Here, \kcache{}-L and \kcache{}-E refer to \kcache{} with lazy and eager rollback schemes respectively.
For workload A, \kcache{}-L shows superior write performance over \kcache{}-E, as it is a write only workload, leading rollback operations to take away bandwidth from actual write operations.
However, both configurations show lower performance in comparison to the write optimized \kcache{} as shown in Figure \ref{fig:eval_avg_throughput}.

Workload B and C present a read-write mix workload, where both rollback schemes achieve similar write throughput, both holding a lead of 36\% and 51\% over ADOC respectively. However, \kcache{}-E shows an increase in read performance, due to rollback allowing more read operations to be performed from Main-LSM, showing that a eager rollback scheme can be more effective for a write/read mixed workload.

\begin{table}[!h]
\scriptsize{}
\centering \footnotesize
\caption{Throughput of range queries for RocksDB, ADOC, and \kcache{} performing workload D.}
\label{tab:eval_range_query}
 \begin{tabular}{|c||c|}
\hline
\cellcolor{gray!30}LSM-KVS & \cellcolor{gray!15}Range Query Throughput (Kops/s) \\ \hline\hline
\cellcolor{gray!15}RocksDB & 302                                                           \\ \hline
\cellcolor{gray!15}ADOC    & 351                                                           \\ \hline
\cellcolor{gray!15}\kcache{} & 100                                                           \\ \hline
\end{tabular}
\end{table}

Table~\ref{tab:eval_range_query} shows the results of range query workloads from workload D. These results prove that \kcache{} is able to fully support the range query operation across the hybrid interfaces. However, \kcache{} still suffers a significant performance hit in comparison to other LSM-KVS. This is in large part due to a lack of read caching mechanism for iterator operations on the Dev-LSM to be accelerated in contrast to the Main-LSM's iterator. 

\subsection{Overhead Analysis}

Through the additional software modules that \kcache{} implements, there are unavoidable overhead processes on top of the core LSM-KVS operations.
A breakdown of all the potential overheads of \kcache{} are covered in Table~\ref{tab:overhead}. 

\begin{table}[!h]
\centering
\footnotesize
\caption{Detailed breakdown of time overheads for \kcache{}'s operations.}
\label{tab:overhead}
\begin{tabular}{|c||c|}
\hline
\cellcolor{gray!30}Operation       & \cellcolor{gray!15}Average Elapsed Time (us) \\ \hline\hline
\cellcolor{gray!15}Detector & 1.37     \\ \hline
\cellcolor{gray!15}Key Insert     & 0.45     \\ \hline
\cellcolor{gray!15}Key Check      & 0.20     \\ \hline
\cellcolor{gray!15}Key Delete     & 0.28     \\ \hline
\end{tabular}
\vspace{-6pt}
\end{table}

The Detector module has the largest overhead impact, with an average of 1.37 microseconds every 0.1 seconds it is used. The Metadata Module is also a required overhead, due to the requirement of maintaining consistency between the dual interfaces. For this there are the key insert, check and delete operations, which on average, takes 0.45, 0.2 and 0.28 microseconds respectively. In practice, during workloads, the largest overhead observed related to the Metadata Manager was the combination of a key check and delete operation, which took 0.48 microseconds.

\subsection{Microscopic Analysis of PCIe Usage}

\begin{figure}[!t]
    \centering
    \includegraphics[width=1.0\linewidth]{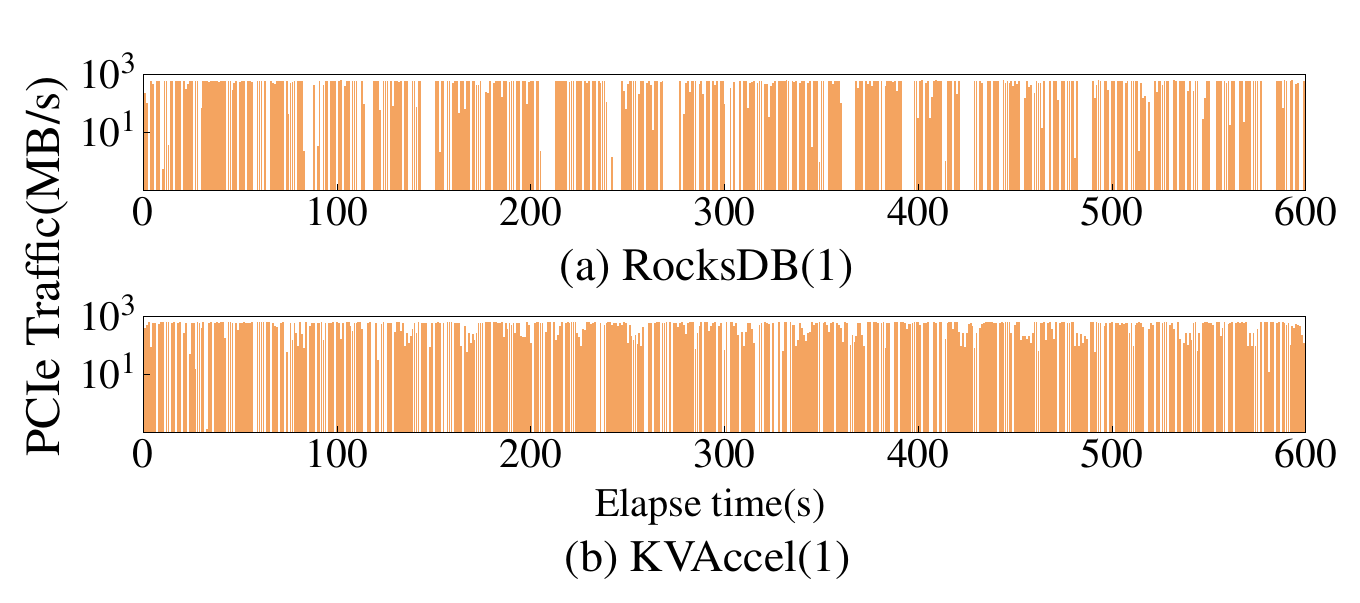}
    \vspace{-16pt}
    \caption{Overview of PCIe bandwidth usage for (a) RocksDB(1) (b) \kcache{}(1) in logarithmic scale.}
    \label{fig:eval_traffic_comp}
    \vspace{-8pt}
\end{figure}

To verify the usage of PCIe bandwidth of \kcache{}, we conducted experiments with Workload A and measured the bandwidth utilization by using Intel PCM~\cite{intelpcm}. Figure~\ref{fig:eval_traffic_comp} shows the results in time series in comparison to baseline RocksDB. It can be observed that \kcache{} takes advantage of its dual interface and demonstrate high PCIe utilization which aligns with the results presented in Figure~\ref{fig:eval_stall_avoid}.

\section{Conclusion}
\label{sec:conc}

There has been extensive research on mitigating write stalls in LSM-tree-based key-value stores. However, these existing studies fall short in overcoming the write stalls and limits the performance gain. 
This study introduces \kcache{}, the first hardware-software co-design that revitalizes the underutilized computational power of SSDs during compaction to avoid write stalls. \kcache{} integrates a dual-interface SSD architecture, dynamically redirecting writes to a key-value interface during host-side write stalls, eliminating the need for complex host-side optimizations, high CPU usage, or additional hardware. We implemented \kcache{} by extending RocksDB to support I/O redirection during write stalls. Our evaluation shows that \kcache{} outperforms ADOC in throughput and CPU efficiency for write-heavy workloads, while both systems perform comparably in mixed read-write scenarios.

\footnotesize{ 
    \bibliographystyle{ieeetr}
    \bibliography{ref} 
}

\end{document}